\documentstyle[12pt]{article}
\renewcommand{\theequation}{\arabic{section}.\arabic{equation}}
\newcommand{\lie}{{\rm Lie}}
\newcommand{\diff}{{\rm Diff}}
\newcommand{\vect}{{\rm vect}}

\begin{document}

\title{Quantum mechanics on Riemannian Manifold in Schwinger's
Quantization Approach II}

\author{Chepilko Nicolai Mikhailovich\\
\small\it
Physics Institute of the Ukrainian Academy of Sciences, 
Kyiv-03 028, Ukraine \\ \small\it e-mail: chepilko@zeos.net
\and
Romanenko Alexander Victorovich\\
\small\it
Kyiv Taras Shevchenko University,
Department of Physics, Kyiv-03 022, Ukraine\\
\small\it e-mail: ar@ups.kiev.ua}

\maketitle

\begin{abstract}

Extended Schwinger's quantization procedure is used for constructing
quantum mechanics on a manifold with a group structure. The considered
manifold $M$ is a homogeneous Riemannian  space with the given action
of isometry transformation group. Using the identification of $M$ with
the quotient space $G/H$, where $H$ is the isotropy group of an
arbitrary fixed point of $M$, we show that quantum mechanics on
$G/H$ possesses a gauge structure, described by the gauge potential
that is the connection 1-form of the principal fiber bundle $G(G/H,
H)$.  The coordinate representation of quantum mechanics and the
procedure for selecting the physical sector of states are developed.

\end{abstract}

\section{Introduction}

The purpose of this paper is to propose the natural development of the
method described in our previous paper \cite{ourpaper}, where we have
introduced an extension of Schwinger's quantization procedure
\cite{schwinger} in order to consider quantum mechanics on a
manifold with a group structure. In \cite{ourpaper} this approach has
been realized for the case of a homogeneous Riemannian manifold
admitting the action of a simply transitive group of isometries.

In this paper we consider a more general type of the $n$-dimensional
homogeneous Riemannian manifold $M$ with the $p$-dimensional group
of isometries acting on $M$ transitively (but not simply transitively).
In this case the part of isometry transformations form an isotropy
group of any point of $M$, so it is possible to treat this
group as a group of local gauge transformations.

The paper is organized as follows. In section (\ref{s1}) we briefly
examine the geometry structure of a homogeneous Riemannian manifold.
Such a manifold is isomorphic to the quotient space $G/H$, where $H$
denotes the isotropy group of an arbitrary fixed point of $M$.
In section (\ref{s2}) the operator Lagrangian $L$ describing a free
particle in the configuration space $G$ is presented in accordance
with \cite{ourpaper}, where its form has been derived by
requiring $L$ to be scalar invariant under a general coordinate
transformation on $G$. The extension of the configuration space from
$M\cong G/H$ to $G$ causes the problem of fixing the gauge
invariance associated with additional degrees of freedom.

To eliminate unphysical states that appears due to the presence of the
gauge degrees of freedom we use $(m+n)$-decomposition that is usual
in the theories of Kaluza-Klein type \cite{kaluza}. After introducing
the special coordinate system on $G$ that provides this decomposition,
the algebra of commutation relations is constructed in section
(\ref{s3}). In section (\ref{s4}) we give the Heisenberg equation of
motion describing a free particle on $G/H$. It turns out that dynamics
on $G/H$ is governed by a Lorentz-type force expressed in terms of the
gauge field by the usual way. The gauge potential is the same as the
connection 1-form of the principal fiber bundle $G(G/H, H)$. In section
(\ref{s5}) the coordinate representation of quantum mechanics on
$G/H$ is discussed. The special feature of the theory consists of the
emergence of the gauge structure induced by some isotropy group
$H\subset G$, and described by the concrete unitary representation of
$H$ in the space of states. Different irreducible representations
determine unequivalent quantum theories on $G/H$. The corresponding
quantum states are classified by eigenvalues of the Casimir
operator of the representation of $H$.

The conclusions obtained from the model we are considering are in
accordance with the conceptions introduced in \cite{tsutsui},
where the method of investigation is somewhat different from ours.

\section{Structure of Homogeneous Riemannian Manifold}
\label{s1}

A smooth manifold $M$ is called a {\it homogeneous} one, if it
admits the transitive action of a Lie group $G$. We assume that
$\dim(M)=n$, $\dim(G)=p>n$. This assumption means that
the action of $G$ on $M$ is not simply transitive; the case of a
simply transitive transformation group (i.e. when $p=n$) has been
investigated in our previous paper \cite{ourpaper}.

A left action of $G$ on $M$ is determined by the following
differentiable map

\begin{equation}
\begin{array}{rcl}
\rho_{g}:&M&\longrightarrow  M\\
&q&\longrightarrow\overline{q}:=\rho_{g}(q)
\end{array}
\label{11}
\end{equation}
where $\rho_{g}\in\diff(M)$ satisfies the conditions

\begin{enumerate}

\item
$
\forall g_{1,2}\in G, \quad
\rho_{g_{1}}\circ \rho_{g_{2}}=\rho_{g_{1}g_{2}};
$

\item
$
\rho_{e}={\rm id}_{M},\quad e\in G \quad \mbox{is the unit element}
$
\end{enumerate}
(here ${\rm id}_{M}$ denotes an identity map on $M$).

For a point $q\in M$ the subset $I(q):=\{g\in G: \rho_{g}q=q\}\subset G$
is a subgroup of $G$. This group is called the
{\it isotropy group} (or {\it stabilizer}) of $q\in M$.
As the action $\rho$ of $G$ is transitive, all isotropy groups are
conjugate, i.e.
$\forall\: \overline{q}=\rho_{g}q$,
$I(\overline{q})=aut_{g} I(q):=g I(q) g^{-1}$.

Fixing an arbitrary point $q_{o}\in M$, we introduce a subgroup
$H:=I(q_{o})\subset G$, $\dim(H)=m$. By means of this subgroup
we construct the quotient space
$G/H=\{gH:g\in G\}$, $\dim(G/H)=p-m=n$.
We denote the element of $G/H$ by $[g]$, where the element $g\in G$
in brackets represents the equivalence class $gH$.

We define a canonical projection

\begin{equation}
\begin{array}{rcl}
p:&G&\longrightarrow G/H\\
  &q&\longrightarrow p(g):=[g]
\end{array}
\label{12}
\end{equation}
that determines the structure of a principal fiber bundle $G(G/H,H)$
with the total space $G$, the base space $G/H$ and the structure group
$H$. The fiber under the point $[g]\in G/H$ is
$p^{-1}([g])=gH\subset G$.

On the other hand, we can naturally define the left transitive action
of $G$ on $G/H$ by the following map

\begin{equation}
\begin{array}{rcl}
\rho_{g}:&G/H&\longrightarrow G/H\\
&[g_{1}]&\longrightarrow\rho_{g}[g_{1}]:=[g\,g_{1}]
\end{array}
\label{13}
\end{equation}
The stabilizer of $[e]\in G/H$ under the transformation (\ref{13})
is the subgroup $H\subset G$, because
$\forall g\in G$, $[gH]=[g]$. For $[g]\in G/H$ we have
$I([g])=g H g^{-1}$.

Hence, there in one-to-one correspondence between the points of
$M$ and those of $G/H$. Namely, $q_{o}\in M$ and $q=\rho_{g}q_{o}$
correspond to $[e]\in G/H$ and $[g]\in G/H$ respectively. We will
identify $M$ with $G/H$ through this text.

In order to analyze the local properties of the principal fiber bundle
$G(G/H, H)$ and the metric structure of $G$ and $G/H$ we introduce
local coordinates on $G$, $H$ and $G/H$.

Hereafter we utilize the notations of indices as follows.
The first lot of Latin capital letters
$A, B,\dots= \overline{1, p}$ is used to represent the frame
$\{\left. T_{A}\right|_{e}: A=\overline{1, p}\}$ of $T_{e} G$
(the space of tangent vectors to $G$ at $e\in G$).
The final lot of Latin capital letters $M, N\dots =\overline{1, p}$
is used to mark the local coordinates $\{x^{M}(g): M=\overline{1, p}\}$
of $g\in G$.

The structure equation for a (left) Lie algebra $\lie(G)$,
which consists of left translations of the elements of $T_{e}G$,
has the usual form

\begin{equation}
[L_{A}, L_{B}]=c^{C}{}_{AB} L_{C},
\label{14}
\end{equation}
where $c^{C}{}_{AB}$ are structure constants of $G$, $L_{A}$ is
a left-invariant vector field over $G$, defined as
$\left. L_{A}\right|_{g}:=d_{e}\,L_{g}(\left. T_{A} \right|_{e})$
($L_{g}$ denotes a left translation map, see Appendix 1).

As far as $H$ is a Lie subgroup of $G$, for the corresponding Lie
algebras we have the similar inclusion: $\lie(H)\subset\lie(G)$
(subalgebra) and $T_{e}H\subset T_{e}G$ (vector subspace).  We can
chose the basic fields of $T_{e}G$ in such a way that part of them
forms the basis of $T_{e}H$. So we can decompose the set $\{T_{A}:
A=\overline{1, p}\}$ as $T_{A}=(T_{a}, T_{i})$, where $\{T_{i}:
i=\overline{n+1, p}\}$ is the basis of $T_{e}H$, marked by small
Latin letters $i, j, k \dots=\overline{n+1, p}$, with the structure
equation

\begin{equation}
[L_{i}, L_{j}]=c^{k}{}_{ij} L_{k}, \quad c^{a}_{ij}=0,
\label{15}
\end{equation}
and $\{T_{a}: a=\overline{1, n}\}$ corresponds to the remaining part
of the basic vectors of $T_{e}G$. In other words, the index
$A=\overline{1, p}$ can be decomposed as $A=(a, i)$.

Due to (\ref{15}) the following system of partial differential
equations

\begin{equation}
L_{i}^{M}\partial_{M}\varphi (x(g))=0
\label{16}
\end{equation}
has $n$ independent solutions
$\{\varphi^{\alpha}(x(g)): \alpha=\overline{1, n}\}$.
Here $L=\{L^{M}_{A}\}$ denotes the matrix of left translations on $G$
(see appendix 1 for its properties).

Now we introduce a new coordinate system on $G$ by means of the
following transformation

\begin{equation}
\left\{
\begin{array}{rcl}
\overline{x}^{\alpha} &=& \varphi^{\alpha}(x(g))\\
\overline{x}^{\mu}  &=& x^{\mu}
\end{array}
\right.
\label{17}
\end{equation}
Hereafter $\{x^{M}\}$  are assumed to refer to the new
coordinate system on $G$. Therefore, the coordinate index
$M=\overline{1, p}$ is decomposed like the group one, namely
$M=(\alpha, \mu)$, where $\alpha,\beta\dots =\overline{1, n}$ and
$\mu,\nu\dots=\overline{n+1, p}$. The meaning of
$\{x^{\alpha}\}$ and $\{x^{\mu}\}$ will be determined later.

In new coordinates the matrix of left translations on $G$
receives the form

\begin{equation}
L=\{L^{M}_{A}\}=
\left(
\begin{array}{cc}
L^{\alpha}_{a} & L^{\mu}_{a}\\
0 & L^{\mu}_{i}
\end{array}
\right)\,,\quad
L^{\alpha}_{i}=0.
\label{18}
\end{equation}
If $\det(L)=\det(L^{\alpha}_{a})\det(L^{\mu}_{i})\ne 0$,
the matrix (\ref{18}) has the inverse

\begin{equation}
L^{-1}:=\overline{L}=\{\overline{L}^{A}_{M}\}=
\left(
\begin{array}{cc}
\overline{L}^{a}_{\alpha} & \overline{L}^{i}_{\alpha}\\
0 & \overline{L}^{i}_{\alpha}
\end{array}
\right)\,,\quad
\overline{L}^{a}_{\mu}=0\,,
\label{19}
\end{equation}
where

\begin{equation}
\overline{L}^{a}_{\alpha}=(L^{-1})^{\alpha}_{a}\,,\quad
\overline{L}^{i}_{\mu}=(L^{-1})^{\mu}_{i}\,,\quad
\overline{L}^{i}_{\alpha}=-\overline{L}^{a}_{\alpha}
L^{\mu}_{a}\overline{L}^{i}_{\mu}
\label{110}
\end{equation}

The matrices $\{L^{a}_{\alpha}\}$, $\{L^{i}_{\mu}\}$ and their inverses
satisfy the equations following from (\ref{14}) and the Maurer-Cartan
equation taking into account the fact that $c^{a}{}_{ij}=0$.

As far as a left Lie algebra commutes with a right one we can generally
write $R^{M}_{A}\partial_{M} L^{N}_{B}=L^{M}_{B}\partial_{M} R^{N}_{A}$
and, in  particular

\begin{equation}
\left\{
\begin{array}{rcl}
\partial_{\mu} R^{\alpha}_{A} &=& 0\\
\partial_{\alpha} R^{\beta}_{A} &=& R^{M}_{A}
\overline{L}^{a}_{\alpha}\partial_{M} L^{\beta}_{a}\,.
\end{array}
\right.
\label{111}
\end{equation}
As to the other relations like (\ref{111}), their explicit forms are
not important in later consideration.

The set of coordinates $\{x^{\alpha}: \alpha=\overline{1, n}\}$
in the decomposition $x^{M}=(x^{\alpha}, x^{\mu})$
is independent on the point at the orbit:
$\forall h\in H$, $x^{\alpha}(gh)=x^{\alpha}(g)$.
To prove this statement one can rewrite $x^{M}(gh)$
as a Tailor expansion
$$
x^{M}(gh)=x^{M}(g)+x^{M}(h)+\dots
$$
taking into account $L^{\alpha}_{i}=0$.

Therefore, we can write the local form of the projection map as

\begin{equation}
\begin{array}{rcl}
p:&G&\longrightarrow G/H\\
&\{x^{\alpha}(g), x^{\mu}(g)\}&\longrightarrow \{x^{\alpha}(g)\}
\label{112}
\end{array}
\end{equation}

Hence, the local coordinate system on $G/H$ can be defined as

\begin{equation}
x^{\alpha}([g]):=x^{\alpha}(g).
\label{113}
\end{equation}

According to (\ref{12}) and (\ref{112}), the action of $G$ on the
quotient space can be presented as follows

\begin{equation}
x^{\alpha}(\rho_{g}[g_{1}])=x^{\alpha}([gg_{1}])=x^{\alpha}(gg_{1}).
\label{114}
\end{equation}

Let $g_{A}(\tau)\in G$, $g_{A}(0)=e$ be the integral curve of the basic
vector field

\begin{equation}
\left. L_{A}\right|_{e}=L^{M}_{A}(e)\left. \partial_{M}\right|_{e}
=\left. T_{A}\right|_{e}\in T_{e}G.
\label{115}
\end{equation}

Then the corresponding vector field over $G/H$, induced by the action of
$G$ on $G/H$ has the form

\begin{equation}
\left. \frac{d}{d\tau}\right|_{\tau=0} x^{\alpha}(g_{A}(\tau)g)=
\left.
\frac{\partial x^{\alpha}(g_{A}(\tau)g)}
{\partial x^{M}(g_{A}(\tau))}
\right|_{g_{A}=e}
\cdot
\left.
\frac{d x^{M}(g_{A}(\tau))}{d \tau}
\right|_{\tau=0}=
R^{\alpha}_{M}(g)\delta^{M}_{A}=R^{\alpha}_{A}(g).
\label{116}
\end{equation}
In particular

$$
\forall h\in H\,, \quad
R^{\alpha}_{A}(h)=R^{\alpha}_{A}(e)=\delta^{\alpha}_{A}
$$
determines the transformations of $I([e])$ on $G/H$, as it must be.

In the special coordinate system, defined in (\ref{17}), using the
diagonal form of the metric of $T_{e}G$ (see appendix 1)

\begin{equation}
\{\eta_{AB}\}=
\left(
\begin{array}{cc}
g_{ab}&0\\
0&g_{ij}
\end{array}
\right).
\label{117}
\end{equation}
we find the following form for the left-invariant metric of $G$

\begin{equation}
\eta_{MN}:=\eta_{AB} \overline{L}^{A}_{M}\overline{L}^{B}_{N}=
g_{ab}\overline{L}^{a}_{M}\overline{L}^{b}_{N}+
g_{ij}\overline{L}^{i}_{M}\overline{L}^{j}_{N};
\label{118}
\end{equation}

\begin{equation}
\eta^{MN}:=\eta^{AB}L^{M}_{A}L^{N}_{B}=
g^{ab}L^{M}_{a}L^{N}_{b}+g^{ij}L^{M}_{i}L^{N}_{j}
\label{119}
\end{equation}

Using the definition of the connection on the principal fiber bundle
$G(G/H, H)$ (see appendix~2)

\begin{equation}
A^{\mu}_{\alpha}:=\overline{L}^{i}_{\alpha}\,L^{\mu}_{i}=
-\overline{L}^{a}_{\alpha}\,L^{\mu}_{a},
\label{120}
\end{equation}
the metric tensor (\ref{118}) and its inverse (\ref{119})
receive the form

\begin{equation}
\{\eta_{MN}\}=
\left(
\begin{array}{cc}
\eta_{\alpha\beta}&\eta_{\mu\beta}\\
\eta_{\alpha\nu}&\eta_{\mu\nu}
\end{array}
\right)=
\left(
\begin{array}{cc}
g_{\alpha\beta}+g_{\rho\sigma}\,A^{\rho}_{\alpha}\,A^{\sigma}_{\beta} &
g_{\mu\rho}\,A^{\rho\beta}\\
A^{\rho}_{\alpha}\,g_{\rho\nu} & g_{\mu\nu}
\end{array}
\right),
\label{121}
\end{equation}

\begin{equation}
\{\eta^{MN}\}=
\left(
\begin{array}{cc}
\eta^{\alpha\beta}&\eta^{\mu\beta}\\
\eta^{\alpha\nu}&\eta^{\mu\nu}
\end{array}
\right)=
\left(
\begin{array}{cc}
g^{\alpha\beta} & -g^{\gamma\beta}\,A^{\mu}_{\gamma}\\
-g^{\alpha\gamma}\,A^{\nu}_{\gamma} &
g^{\mu\nu}+g^{\gamma\delta}\,A^{\mu}_{\gamma}\,A^{\nu}_{\delta}
\end{array}
\right),
\label{122}
\end{equation}
where we denote the metric tensors of $G/H$ and $H$ as

\begin{equation}
g_{\alpha\beta}=
g_{ab}\overline{L}^{a}_{\alpha}\overline{L}^{b}_{\beta}\,,\quad
g^{\alpha\beta}=g^{ab}L^{\alpha}_{a}L^{\beta}_{b}
\label{123}
\end{equation}
and

\begin{equation}
g_{\mu\nu}=
g_{ij}\overline{L}^{i}_{\mu}\overline{L}^{j}_{\nu}\,,\quad
g^{\mu\nu}=g^{ij}L^{\mu}_{i}L^{\nu}_{j}
\label{124}
\end{equation}
correspondingly.

Note that the similar form of the metric tensor (\ref{121}), (\ref{122})
appears in the Kaluza-Klein scheme \cite{kaluza}.

The right-invariant vector fields $R_{A}=R_{A}^{M}\partial_{M}$
are the Killing vectors for the left-invariant metric $\eta_{MN}$ of
$G$, as it follows from the definition of $\{R^{M}_{A}\}$.

Using (\ref{121}) we can rewrite the Killing equation for
$\{\eta_{MN}\}$ in the following form

\begin{equation}
-R^{\gamma}_{A}\partial_{\gamma} g^{\alpha\beta}+
g^{\alpha\gamma}\partial_{\gamma} R^{\beta}_{A}+
g^{\beta\gamma}\partial_{\gamma} R^{\alpha}_{A}=
R^{\mu}\partial_{\mu} g^{\alpha\beta}.
\label{125}
\end{equation}
On the other hand, we can define the metric of $G/H$ from the
metric $\{\eta_{MN}\}$ of $G$, requiring the map $d\,p$ to be
an isometry.

A direct calculation shows, that the metric of $G/H$ is
described by $\{g_{\alpha\beta}\}$. If we treat $G/H$ as a
homogeneous Riemannian manifold, we have to conclude from (\ref{125})
that $\partial_{\mu} g^{\alpha\beta}=0$. This condition means that the
Lie variation of $\eta_{MN}$ for left-invariant vector fields
$L_{A}=L^{M}_{A}\partial_{M}$ vanishes, i.e. $\{\eta_{MN}\}$ is both
invariant. It is possible if and only if $G$ is semisimple group.
In this case the metric $\{\eta_{AB}\}$ of $T_{e}G$ can be identified
with an adjoint-invariant Cartan-Killing form.

It is useful to write down here the explicit form of the Killing
equations $\delta\eta_{MN}=0$ for left-invariant vector fields
in the special coordinate system

\begin{eqnarray}
&&\partial_{\mu}\, g_{\alpha\beta}=0\nonumber \\
&&L^{\mu}_{i}\,\partial_{\mu}A^{\nu}_{\alpha}=
A^{\mu}_{\alpha}\partial_{\mu}L^{\nu}_{i}-\partial_{\alpha}L^{N}_{i}\\
&&L^{\sigma}_{i}\,\partial_{\sigma} g_{\mu\nu}+
g_{\mu\sigma}\partial_{\nu}L^{\sigma}_{i}+
g_{\nu\sigma}\partial_{\mu}L^{\sigma}_{i}=0.\nonumber \label{125a}
\end{eqnarray}

Hence, we have shown that the homogeneous Riemannian manifold $M$ can
be identified with the quotient space $G/H$, where $G$ is the isometry
group of $M$ and $H=I(q_{o})$ is a stabilizer of an arbitrary fixed
point $q_{o}\in M$. The general geometric analyses of this problem is
presented in \cite{kobayashi}.

\section{Quantum Lagrangian for Free Particle in Homogeneous
Riemannian Space}
\label{s2}
\setcounter{equation}{0}

Now we consider quantum mechanics for a free particle in a
homogeneous Riemannian  space $M$, $\dim(M)=n$ with a given action
of the group $G$ of isometries, $\dim(G)=p>n$.
As it has been shown in the previous section, the configuration
space $M$ can be identified with the quotient space $G/H$, where
$H$, $\dim(H)=m=p-n$ denotes the stabilizer of an arbitrary fixed
point of $M$.

In the construction of quantum theory, based on Schwinger's
quantization approach, the key role is played by the realization
of a Lie algebra $\lie(G)$ of $G$, induced by the realization of
$G$ on $M$. The set of independent Killing vectors, associated
with the basis of $\lie(G)$, possesses the properties of
permissible variations $\{\delta q^{\mu}\}$, where $\{q^{\mu}\}$
denote the set of coordinate operators describing the position of
a particle.

In the case of the homogeneous Riemannian manifold $G/H$ the
number $p=\dim(G)$ of independent Killing vectors is bigger than
the dimension $n$ of the manifold $G/H$. Some of these vectors  form
the representation of the isotropy group $I([g])$, $\dim(I([g]))=m=p-n$
that depends on a choice of a point $[g]\in G/H$. The transformations
of $I([g])$ we can treat as local gauge ones. Therefore, we can divide
independent Killing vectors at $[g]\in G/H$ into two sets. The first
one generates non-trivial transformations on $G/H$, the second one
realizes the action of the stabilizer subgroup $I([g])\subset G$ on $M$
(the group of local gauge transformations).

Making an attempt to realize Schwinger's quantization procedure
immediately as in \cite{ourpaper}, one observe the same difficulties as in
usual gauge models. Here, as in any theory with first-class
constraints, there is a problem of fixing the gauge degrees of freedom,
which number in our model equals to $m=\dim(G)-\dim(G/H)$.

This procedure is performed in the present model by means of
introducing a new configuration space $G$ (the space of an isometry
group). The local coordinate system in $G$ is described by

\begin{equation}
x^{M}=\{x^{M}(g): x^{M}(g)=\{x^{\alpha}([g]), x^{\mu}(g)\}, g\in G\}\,,
\quad
M=\overline{1, p},\:\alpha=\overline{1, n},\:\mu=\overline{n+1, p}
\label{21}
\end{equation}
where $\{x^{\mu}(\mbox{ })\}$ are coordinates in the orbit $gH$.

The metric $\{\eta_{MN}(g): M,N=\overline{1, p}\}$ has been
defined in the previous section by the formulae (\ref{121}),
(\ref{122}).

Of course, the enlargement of the number of degrees of freedom from
$n=\dim(G/H)$ to $p=\dim(G)$ brings the appearance of unphysical quantum
states. The procedure for its elimination will be presented later.

The quantum Lagrangian describing a free particle in the new
configuration space $G$ has the following form

\begin{equation}
L_{G}:=\frac{1}{2} \dot{x}^{M} \eta_{MN}(g) \dot{x}^{N}-
U_{q}(g)
\label{22}
\end{equation}
which has been introduced in \cite{ourpaper}. Here $U_{q}(g)$ denotes a
so-called ``quantum potential''. Its role consists of providing the
scalar invariance of (\ref{22}) under a general coordinate
transformation $x^{M}\to \overline{x}^{M}=\overline{x}^{M}(x)$ on $G$
(note that $[x^{M}, \dot{x}^{N}]\ne 0$).

Taking into account $(m+n)$-decomposition of the metric in a
special coordinate system, introduced in the previous section, the
Lagrangian (\ref{22}) can be written as

\begin{equation}
L_{G}=\frac{1}{2}\dot{x}^{\alpha} g_{\alpha\beta}([g]) \dot{x}^{\beta}+
\frac{1}{2}
\left( \dot{x}^{\mu}+\dot{x}^{\alpha}A^{\mu}_{\alpha}(g) \right)
g_{\mu\nu}(g)
\left( \dot{x}^{\nu}+A^{\nu}_{\beta}(g)\dot{x}^{\beta} \right)
\label{23}
\end{equation}
where the first term corresponds to the kinetic energy of a particle on
$G/H$, while the appearance of the second term is caused by the
extension of the physical configuration space from $G/H$ to $G$.

Since $G$ acts on itself simply transitively, the method of constructing
quantum theory based on the Lagrangian (\ref{22}) (or (\ref{23}))
coincides with that of developed in \cite{ourpaper}.
Note that due to semisimplicity of $G$, there are two equivalent sets
of Killing vectors $\{v^{M}_{A}(g)\}$, that correspond to the matrices
of left and right translations on $G$, i.e. $\{L^{M}_{A}(g)\}$ and
$\{R^{M}_{A}(g)\}$ correspondingly.

In accordance with \cite{ourpaper} the permissible variations of the
coordinate operators $\{x^{M}\}$ can be written in the following form

\begin{equation}
\delta x^{M}=\varepsilon^{A} v_{A}^{M}(g), \quad G=\varepsilon^{A}
v^{M}_{A}\circ p_{M}, \quad p_{M}:=\eta_{MN}\circ \dot{x}^{N}.
\label{24-5}
\end{equation}
Here $\varepsilon^{A}$ is an infinitesimal $c$-number parameter of a
coordinate transformation on $G$.

\section{Algebra of Commutation Relations}
\label{s3}
\setcounter{equation}{0}

Constructing the algebra of commutation relations for
operators describing quantum mechanics of the particle on $G/H$, we
will use the fact that the algebra to be found is contained in the
wider one associated with quantum mechanics on $G$.

At first we consider the right isometries of the metric
$\{\eta_{MN}\}$.
In this case, in accordance with the results of the previous sections,
Killing vectors coincide with left-invariant vector fields
$\{ L^{\mu}_{i}(g)\left. \partial_{\mu}\right|_{g}\}$, which determine
the generator of right translations as in following

\begin{equation}
G_{i}=L^{M}_{i}\circ p_{\mu}\,,\quad
p_{\mu}:=\eta_{\mu N}\circ \dot{x}^{N}=g_{\mu\nu}\circ
\left( \dot{x}^{\nu}+A^{\nu}_{\alpha}\circ \dot{x}^{\alpha} \right)
\label{31}
\end{equation}
In this case

\begin{equation}
\begin{array}{rcl}
\delta_{i} x^{\alpha}&=0&=\displaystyle\frac{1}{i\hbar}
[x^{\alpha}, L^{\mu}_{i}\circ p_{\mu}] \\[3mm] \delta_{i}x^{\mu}&=
L^{\mu}_{i}&=\displaystyle\frac{1}{i\hbar} [x^{\mu},
L^{\mu}_{i}\circ p_{\mu}]\,.
\end{array}
\label{32}
\end{equation}

Since $[x^{M}, L^{\mu}_{i}]=0$ and $\det(L^{\mu}_{i})\ne 0$
we can conclude from (\ref{32}) that

\begin{equation}
\left[ x^{M}, p_{\mu} \right]=i\hbar \delta^{M}_{\mu}.
\label{33}
\end{equation}

For the case of an arbitrary function $f(\{x^{M}\})$ depending on the
coordinates $\{x^{M}\}$ we have

\begin{equation}
\delta_{i} f=L^{M}_{i}\partial_{M} f=L^{\mu}_{i}\partial_{i} f=
\frac{1}{i\hbar}\left[ f, L^{\nu}_{i}\circ p_{\nu} \right]\,,
\label{34}
\end{equation}
then

\begin{equation}
\left[ f, p_{\mu} \right]=i\hbar \partial_{\mu} f\,.
\label{35}
\end{equation}
Taking into account the transformation law of $p_{\mu}$ under the
transformation $x^{M}\to x^{M}+\delta x^{M}$ one can easily find

\begin{equation}
[p_{\mu}, p_{\nu}]=0.
\label{36}
\end{equation}
Developed by this way commutation relations determine quantum
mechanics on the orbit $gH$.

Using the commutation relations (\ref{33}-\ref{36}) and the structure
equation for $\{\L^{\mu}_{i}\}$ one can directly prove that

\begin{equation}
[p_{i}, p_{j}]=-i\hbar c^{k}{}_{ij} p_{k}
\label{37}
\end{equation}
where $p_{i}:=L^{\mu}_{i}\circ p_{\mu}$.

Further we consider the left isometries of the metric $\{\eta_{MN}\}$
which are described by the set of the Killing vectors
$\{R^{M}_{A}\partial_{M}\}$. The generator of these transformations has
the form

\begin{equation}
G_{A}=R^{M}_{A}\circ p_{M}=R^{\alpha}_{A}\circ \pi_{\alpha}+
\left( R^{\mu}_{A}+A^{\mu}_{\alpha} R^{\alpha}_{A} \right)\circ p_{\mu}
\label{38}
\end{equation}
where we denote

\begin{eqnarray}
p_{\alpha}&=&\eta_{\alpha M}\circ \dot{x}^{M}=
\pi_{\alpha}+A^{\mu}_{\alpha}\circ p_{\mu}\,,\quad
\pi_{\alpha}=g_{\alpha\beta} \circ \dot{x}^{\beta}\,,\nonumber \\
p_{\mu}&=&g_{\mu\nu}\circ
\left( \dot{x}^{\nu}+A^{\nu}_{\alpha} \circ \dot{x}^{\alpha} \right).
\label{39}
\end{eqnarray}
The vector $\{\pi_{\alpha}\}$ rewritten as

\begin{equation}
\pi_{\alpha}=p_{\alpha}-A^{i}_{\alpha}\circ p_{i}\,,\quad
A^{i}_{\mu}=\overline{L}^{i}_{\mu} A^{\mu}_{\alpha}=
\overline{L}^{i}_{\alpha}
\label{310}
\end{equation}
has the sense of the momentum operator of a free particle on $G/H$.
As far as $\{R^{\alpha}_{A}\partial_{\alpha}\}$ are Killing vectors of
$G/H$, the first term in (\ref{38}) coincides with the generator of
isometry transformations of the metric of $G/H$, while the second one
is caused by extention of the configuration space from $G/H$ to $G$.

To define the operator properties of $\pi_{\alpha}$, we consider
the variation of an arbitrary function $f(x)$ of only
$\{x^{M}\}$'s:

\begin{equation}
\delta_{A} f=R^{M}_{A}\partial_{M} f=\frac{1}{i\hbar}[f, G_{A}].
\label{311}
\end{equation}

Substituting the explicit form of the generator $G$ into (\ref{311})
one can rewrite (\ref{311}) as

\begin{equation}
R^{\alpha}_{M}\circ [f, \pi_{\alpha}]=
i\hbar R^{\alpha}_{A}\left( \partial_{\alpha}-A^{i}_{\alpha} L_{i}
\right) f\,,\quad
L_{i}=L^{\mu}_{i}\partial_{i}\,.
\label{312}
\end{equation}
Using the fact that $\{\overline{R}^{A}_{M}\}$ is the inverse of
$\{R^{\mu}_{A}\}$ we can contract (\ref{312}) with
$\overline{R}^{A}_{M}$, then

\begin{equation}
[f, \pi_{\alpha}]=i\hbar D_{\alpha} f\,,\quad
D_{\alpha}:=\partial_{\alpha}-A^{i}_{\alpha} L_{i}\,.
\label{313}
\end{equation}
The commutator of new derivatives $D_{\alpha}$ acts on a scalar
function $f(\{x^{M}\})$ as

\begin{equation}
\left[ D_{\alpha}, D_{\beta} \right] f=-F^{i}{}_{\alpha\beta}
L_{i}
\label{314}
\end{equation}
where

\begin{equation}
F^{i}{}_{\alpha\beta}=\partial_{\alpha} A^{i}_{\beta}-
\partial_{\beta} A^{i}_{\alpha}+
c^{i}{}_{jk}\,A^{j}_{\alpha}\,A^{k}_{\beta}\,.
\label{315}
\end{equation}

A direct calculation leads to

\begin{equation}
\left[ \pi_{\alpha}, \pi_{\beta} \right]=
i\hbar F^{i}{}_{\alpha\beta}\circ p_{i}\,.
\label{316}
\end{equation}

The object $\{F^{i}{}_{\alpha\beta}\}$ in (\ref{314}-\ref{316})
can be treated as a strength tensor of the gauge field $A^{i}_{\alpha}$
on $G/H$. The corresponding gauge group is the isotropy group
$H\subset G$.

\section{Equations of Motion and Hamiltonian}
\label{s4}
\setcounter{equation}{0}

Using the same procedure as developed in  \cite{ourpaper} one can
construct the Hamiltonian for the system in the configuration space
$G$, expressed in terms of momentum operators $p_{M}=\eta_{MN}\circ
\dot{x}^{N}$. The Hamiltonian is completely determined by the initial
Lagrangian

\begin{equation}
H_{G}=\frac{1}{2}\left( p_{M}-\frac{i\hbar}{2} \Gamma_{M} \right)
\eta^{MN} \left( p_{N}+\frac{i\hbar}{2}\Gamma_{N} \right)=
\frac{1}{2}p_{M}\eta^{MN} p_{N}+V_{G}(x)
\label{41}
\end{equation}
where

\begin{equation}
V_{G}=\frac{\hbar^{2}}{4}\left( \partial_{M} \Gamma^{M}+
\frac{1}{2}\Gamma_{M}\Gamma^{M}  \right)\,,\quad
\Gamma_{M}=\Gamma^{N}{}_{MN}=\frac{1}{2\det(\eta_{MN})}\partial_{M}
\det(\eta_{MN})\,.
\label{42}
\end{equation}

Here $\Gamma^{M}{}_{N_{1} N_{2}}$ denotes the Christoffel symbol
constructed with the metric $\{\eta_{MN}\}$.

Using the properties of $(m+n)$-decomposition one can rewrite
(\ref{41}) as

\begin{equation}
H_{G}=H_{G/H}+H_{orb}\,,
\label{43}
\end{equation}
where

\begin{equation}
H_{G/H}=\frac{1}{2}
\left( \pi_{\alpha}-\frac{i\hbar}{2}\Gamma_{\alpha} \right)
g^{\alpha\beta}
\left( \pi_{\beta}+\frac{i\hbar}{2}\Gamma_{\beta} \right)=
\frac{1}{2}\pi_{\alpha} g^{\alpha\beta} \pi_{\beta}+V_{G/H}\,,
\label{44}
\end{equation}

\begin{equation}
H_{orb}=\frac{1}{2} p_{i} g^{ij} p_{j}=\frac{1}{2} \left(
p_{\mu}-\frac{i\hbar}{2}\Gamma_{\mu} \right) g^{\mu\nu} \left(
p_{\nu}+\frac{i\hbar}{2}\Gamma_{\nu} \right)=
\frac{1}{2}p_{\mu}g^{\mu\nu}p_{\nu}+V_{orb}
\label{45}
\end{equation}
The objects $\Gamma_{\alpha}=\Gamma^{\beta}{}_{\alpha\beta}$
and $\Gamma_{\mu}=\Gamma^{\nu}{}_{\mu\nu}$ are defined analogously to
$\Gamma_{M}$. The ``quantum potentials'' $V_{G/H}$ and $V_{orb}$
have the form

\begin{equation}
V_{G/H}=\frac{\hbar^{2}}{4}\left( \partial_{\alpha}\Gamma^{\alpha}+
\frac{1}{2}\Gamma^{\alpha}\Gamma_{\alpha} \right)\,,\quad
V_{orb}=\frac{\hbar^{2}}{4}\left( \partial_{\mu}\Gamma^{\mu}+
\frac{1}{2}\Gamma^{\mu}\Gamma_{\mu} \right)\,.
\label{46}
\end{equation}

The Heisenberg equations of motion describing dynamics in $G$ can be
derived by means of $(m+n)$-decomposition of the metric

\begin{equation}
\dot{p}_{M}=\frac{1}{i\hbar}\left[ p_{M}, H_{G} \right]\,.
\label{47}
\end{equation}
Performing the direct calculation and using commutative relations, one
can find

\begin{equation}
\dot{\pi}_{\alpha}=
\frac{1}{2}\pi_{\beta} \partial_{\alpha} g^{\beta\gamma} \pi_{\gamma}+
\left( F_{\alpha\beta}\circ g^{\beta\gamma} \right) \circ \pi_{\gamma}+
\partial_{\alpha} V_{G/H}\,,
\label{48}
\end{equation}

\begin{equation}
\dot{p}_{i}=0\,,
\label{49}
\end{equation}
where we denote $F_{\alpha\beta}:=F^{i}{}_{\alpha\beta}\circ p_{i}$.

So we can conclude from (\ref{48}, \ref{49}) that the motion of a
particle is governed by a Lorentz-type force represented by the
second term in the right hand side of (\ref{48}). This object is
determined by the strength tensor (\ref{315}) of the gauge potential
$A^{i}_{\alpha}$. The motion of a particle on the orbit $gH$ is
completely flat due to the conservation law (\ref{49}).

The main result of this section consists of the emergence of a gauge
structure in quantum theory on a homogeneous manifold. Such a structure
is induced by additional degrees of freedom caused by an isotropy
group. This result is not surprising because the given theory can be
considered as a version of the Kaluza-Klein scheme, that has been
exhaustively investigated in a great number of works \cite{kaluza}.

\section{Coordinate Representation and Physical Sector of Sta\-tes}
\label{s5}
\setcounter{equation}{0}

The procedure for constructing the quantum space of states for
quantum mechanics on a homogeneous Riemannian manifold with the
simply transitive action of the transformation group $G$ has been
introduced in \cite{ourpaper}. The obtained results are quite
applicable in the case we are considering. The problem arising here is
how to eliminate unphysical states (that does not refer to quantum
mechanics on $G/H$) from the whole set of states describing quantum
mechanics on $G$. The simplest way to perform such a procedure consists
of using $(m+n)$-decomposition.

According to \cite{ourpaper} the coordinate representation of the
operators, corresponding to quantum mechanics on $G$, is defined by
its action on the wave functions

\begin{equation}
\psi(x):=\left\langle x \right.\left| \psi \right\rangle\,.
\label{51}
\end{equation}
(here $\left| x \right\rangle$ is an eigenvector of the coordinate
operator and $\left| \psi \right\rangle$ denotes an arbitrary vector of
state)
has the following form

\begin{equation}
\begin{array}{rcl}
\hat{x}^{M}&=&x^{M}\\ \hat{p}_{M}&=&-i\hbar\left( \partial_{M}+
\displaystyle\frac{1}{2}\,\Gamma_{M} \right)\,.
\end{array}
\label{52}
\end{equation}
Similarly we can write

\begin{equation}
\hat{H}_{G}=-\frac{\hbar^{2}}{2}\left(\partial_{M}+\Gamma_{M}\right)
\eta^{MN} \left(\partial_{M}+\Gamma_{M}\right)=
-\frac{\hbar^{2}}{2}\frac{1}{\sqrt{\eta}}\partial_{M} \left(
\sqrt{\eta}\,\eta^{MN}\,\partial_{N}  \right)\,, \label{53}
\end{equation}
where $\eta=\det(\eta_{MN})$.

The wave function (\ref{51}) satisfies the Schr\"odinger equation

\begin{equation}
-\frac{\hbar^{2}}{2}\frac{1}{\sqrt{\eta}}
\partial_{M}\left( \sqrt{\eta}\,\eta^{MN}\partial_{N}\psi
\right)=E\psi\,.
\label{54}
\end{equation}
The coordinate representation of the generator of permissible
variations on $G$ has the form

\begin{equation}
\hat{G}=\varepsilon^{A}\hat{v}^{M}_{A}\circ \hat{p}_{M}=
-i\hbar\varepsilon^{A} v^{M}_{A}\partial_{M}\,,
\label{55}
\end{equation}
that can be derived using (\ref{52}) and the properties of the Killing
vectors $\{v^{M}_{A}\partial_{M}: a=\overline{1, p}\}$.

Hence, the coordinate and momentum operators can be rewritten in terms
of $(m+n)$-decomposition as

\begin{equation}
\hat{x}^{\mu}=x^{\mu}\,,\quad
\hat{p}_{\mu}=-i\hbar\left(
\partial_{\mu}+\frac{1}{2}\Gamma_{\mu}\right)
\label{56}
\end{equation}
for the operators describing quantum mechanics on the orbit $gH$,
and

\begin{equation}
\hat{x}^{\alpha}=x^{\alpha}\,,\quad
\hat{p}_{\alpha}=-i\hbar\left(
\partial_{\alpha}+\frac{1}{2}\Gamma_{\alpha}\right)
\label{57}
\end{equation}
for the operators describing quantum mechanics on the quotient
space $G/H$.

Further we consider the procedure of selection of the physical sector,
or, in other words, the states represented by the subset
$L_{2}(G/H)\subset L_{2}(G)$ that describe quantum mechanics on
$G/H$.

The wave function $\psi\in L_{2}(G)$ performs the map

$$
\psi: G\longrightarrow {\bf C}^{n}\,,
$$
that can be restricted to the function on $G/H$ by means of the section
of $G(G/H, H)$

\begin{equation}
\begin{array}{rcl}
s: & G/H &\longrightarrow G\\
   & [g] &\longrightarrow s([g])\in gH
\end{array}
\label{58}
\end{equation}
that meets the condition $\rho(s([g]))=g$ and performs the
correspondence between the equivalence class $[g]$ and its
representative $gh=s([g])\in gH\subset G$ for some $h\in H$.
In this expression the element $h\in H$ completely determines the
section s, therefore we denote this section as $s_{h}$.

Hence

\begin{equation}
\begin{array}{rcl}
\psi\circ s_{h}:=\phi: &G/H& \longrightarrow  {\bf C}^{n} \\
                       &[g]& \longrightarrow  \phi([g])=\psi(s_{h}([g]))\equiv
\psi(gh)
\end{array}
\label{59}
\end{equation}
is the wave function on $G$.

The matrix elements of physical observables calculated on the wave
functions $\phi=\psi\circ s\in L_{2}(G/H)$ have to be independent on
the choice of the section $s_{h}$. It is possible if and only if the
wave functions $\phi:=\psi\circ s_{h}$ and $\phi':=\psi\circ s_{h'}$
are connected by the unitary transformation

\begin{equation}
\phi'([g])\equiv \psi(gh')\equiv \psi(g h h^{-1} h')=
U(h, h')\psi(gh)=U(h, h')\phi([g])\,.
\label{510}
\end{equation}
Therefore one can show that the wave functions of the physical sector
obey the condition

\begin{equation}
\psi_{phys}(gh)=\sigma_{h^{-1}} \psi_{phys}(g)\,.
\label{511}
\end{equation}
where $\sigma_{h^{-1}}$ is the right unitary representation of
$H\subset G$ in ${\bf C}^{n}$ (while $\sigma_{h}$ is the left one).

The representation of $H$ on physical states induces the representation
of $\lie(H)$ as

\begin{equation}
\tilde{\sigma}_{i}:=\left. \frac{d}{d\tau}\right|_{\tau=0}
\sigma_{h_{i}(\tau)}\in\vect({\bf C}^{n})\cong {\bf C}^{n}\,,
\label{512}
\end{equation}
where $h_{i}(\tau)\in H$ is an integral curve for the basic element
$T_{i}|_{e}\in T_{e} H$. The connection between (\ref{512}) and the
generator of coordinate transformation can be expressed as

$$
\tilde{\sigma}_{i}\psi_{phys}=\left.\frac{d}{d\tau}\right|_{\tau=0}
\psi_{phys}(g h_{i}^{-1}(\tau))=
\frac{\partial \psi_{phys}(g h_{i}^{-1})}
{\partial x^{\alpha}(g h_{i}^{-1})}
\frac{\partial x^{\alpha}(g h_{i}^{-1})}{\partial x^{\mu}(h_{i}^{-1})}
\left.\frac{d x^{\mu}(h_{i}^{-1})}{d\tau}\right|_{\tau=0}=
-L^{\alpha}_{i}(g) \partial_{\alpha}\psi_{phys}\,.
$$
Hence, the wave functions of the physical sector of states satisfy the
equation

\begin{equation}
L^{\alpha}_{i}\partial_{\alpha}\psi_{phys}\equiv
\frac{1}{i\hbar}\hat{p}_{i}\psi_{phys}=
-\tilde{\sigma}_{i}\psi_{phys}\,,
\label{513}
\end{equation}
that points to the fact that $p_{i}=L^{\mu}_{i}\circ p_{\mu}$
describe the representation of $\lie(H)$.

Using (\ref{513}) one can express the ``horizontal derivative''
$D_{\alpha}$ in terms of the generators $\tilde{\sigma}_{i}$.
The derivative $D_{\alpha}$ acts on physical sector as

\begin{equation}
D_{\alpha}=\left( \partial_{\alpha}+A^{i}_{\alpha}
\tilde{\sigma}_{i}  \right)\psi_{phys}\,.
\label{514}
\end{equation}
This formula coincides with the definition of the invariant derivative
associated with the action of the gauge group.

Finally, using (\ref{514}) in (\ref{53}) we find a coordinate
representation of $\hat{H}_{G/H}$ on physical sector:

\begin{equation}
\hat{H}_{G/H}=-\frac{\hbar^{2}}{2}
\left( D_{\alpha}+\Gamma_{\alpha} \right)g^{\alpha\beta}
D_{\beta}-\frac{\hbar^{2}}{2}\hat{C}
\label{515}
\end{equation}
where $\hat{C}=\eta^{ij}\tilde{\sigma}_{i}\tilde{\sigma}_{j}$
is the Casimir operator of the unitary representation of $H$.

According to general theory of unitary representations \cite{groups},
the irreducible representations of Lie groups are finite dimensional
and can be described by eigenvalues of the Casimir operator.

Therefore, the given irreducible unitary representation describes one
from several unequivalent theories on $G/H$ based on the Hamiltonian
(\ref{53}).

The analyses of unequivalent quantum theories has been performed
in \cite{tsutsui} in terms of representation theory of Weyl
relations. Our final results, obtained by means of extended
Schwinger quantization scheme, completely correspond to the
results of (\cite{tsutsui}).

\section{Summary and Discussion}

Quantum mechanics on the homogeneous manifold $G/H$ has been
constructed using our extension of Schwinger quantization procedure.
The essential feature of quantum mechanics on $G/H$ consists of the
appearance of the gauge structure induced by some unitary (irreducible)
representation of the isotropy subgroup $H\subset G$ ($H$ plays the
role of a gauge group). The gauge field corresponds to the connection
1-form of the fiber bundle $G(G/H, H)$. There exist a number of
unequivalent quantum theories classified by eigenvalues of the
Casimir operator of the unitary representation.

Successful development of quantum mechanics on a homogeneous Riemannian
manifold with simply and non-simply transitive transformation groups of
isometries shows that extended Schwinger's quantization scheme
is suitable in constructing quantum mechanics on a manifold with a
group structure. This approach can be applied to analyze a number of
models such as Kaluza-Klein theories \cite{kaluza} or to generalize
simplest hadron models \cite{hadron}.

\section*{Appendix 1. Realizations of Lie Groups on Manifolds}
\setcounter{equation}{0}
\renewcommand{\theequation}{A1.\arabic{equation}}

\subsection*{Lie Groups and Lie Algebras}

Let $G$ be a $p$-dimensional Lie group with a local coordinate system
$\{x^{M}(g): g\in G, M=\overline{1, p}\}$ at a point $g\in G$.

Left- and right translations are defined as

\begin{equation}
\begin{array}{rcl}
L_{g}: &G\longrightarrow& G\\
       &h\longrightarrow& L_{g} h:=gh\,,
\end{array}
\qquad
\begin{array}{rcl}
R_{g}: &G\longrightarrow& G\\
       &h\longrightarrow& R_{g} h:=hg\,.
\end{array}
\label{a11}
\end{equation}
In the tangent space $T_{h} G$, $L_{g}$ and $R_{g}$ induce the
following differential maps

\begin{equation}
d\,L_{g}: T_{h} G\longrightarrow  T_{gh} G\,,\quad
d\,R_{g}: T_{h} G\longrightarrow  T_{hg} G\,.
\label{a12}
\end{equation}
In the local coordinate system $\{x^{M}(\cdot)\}$ the element
$\left. A\right|_{h}\in T_{h} G$ can be written as

\begin{equation}
\left.A\right|_{h}=a^{M}(h) \left.\partial_{M} \right|_{h}\,.
\label{a13}
\end{equation}
Therefore we can express the transformations (\ref{a12}) in the local
form

\begin{equation}
d\,L_{g}(A|_{h})=a^{M}(h)
\frac{\partial x^{N}(gh)}{\partial x^{M}(h)}
\frac{\partial }{\partial x^{N}(gh)}=
a^{M}(h) L^{N}_{M}(gh, h)\frac{\partial }{\partial x^{N}(gh)}\,,
\label{a14}
\end{equation}

\begin{equation}
d\,R_{g}(A|_{h})=a^{M}(h)
\frac{\partial x^{N}(hg)}{\partial x^{M}(h)}
\frac{\partial }{\partial x^{N}(hg)}=
a^{M}(h) R^{N}_{M}(hg, h)\frac{\partial }{\partial x^{N}(hg)}\,,
\label{a15}
\end{equation}
where we denote the matrices of left and right translations together with
their inverses as

\begin{equation}
L^{N}_{M}(gh, h)=\frac{\partial x^{N}(gh)}{\partial x^{M}(h)}\,,\quad
\overline{L}^{M}_{N}(h, gh)=
\frac{\partial x^{M}(h)}{\partial x^{N}(gh)}\,,
\label{a16}
\end{equation}

\begin{equation}
R^{N}_{M}(hg, h)=\frac{\partial x^{N}(hg)}{\partial
x^{M}(h)}\,,\quad \overline{R}^{M}_{N}(h, hg)= \frac{\partial
x^{M}(h)}{\partial x^{N}(hg)}\,,
\label{a16a}
\end{equation}
(here $\overline{L}=L^{-1}$, $\overline{R}=R^{-1}$)

The Lie algebras of left- and right-invariant vector fields are
constructed by left and right translations of the elements of $T_{e}G$
($e\in G$ denotes the unit element) correspondingly.

We denote the basic element of $T_{e}G$ as

\begin{equation}
T_{A}|_{e}=\frac{\partial }{\partial x^{A}(e)}:=\partial_{A}|_{e} \,.
\label{a18}
\end{equation}
Then the set of left [right] invariant vector fields

\begin{equation}
L_{A}|_{g}:=d\,L_{g}
\left( \left.\frac{\partial }{\partial x^{A}}\right|_{e} \right)=
d\,L_{g}(T_{A}|_{e})=L^{M}_{A}(g)
\left. \frac{\partial }{\partial x^{M}} \right|_{g}
\label{a19}
\end{equation}

\begin{equation}
\left[
R_{A}|_{g}:=d\,R_{g}
\left( \left.\frac{\partial }{\partial x^{A}}\right|_{e} \right)=
d\,R_{g}(T_{A}|_{e})=R^{M}_{A}(g)
\left. \frac{\partial }{\partial x^{M}} \right|_{g}
\right]
\label{a19r}
\end{equation}
form a basis of the left [right] Lie algebra $\lie(G)$ of the Lie
group $G$.  The corresponding matrices $L(g)=\{L^{M}_{A}(g)\}$ and
$R(g)=\{R^{M}_{A}(g)\}$ are obtained by the reduction of
(\ref{a16}), (\ref{a16a}):

\begin{equation}
L^{M}_{A}(g)=L^{M}_{A}(gh, h)|_{h=e}\,,\quad
\overline{L}^{A}_{M}(g)=\overline{L}^{A}_{M}(h, gh)|_{h=e}\,,
\label{a110}
\end{equation}

\begin{equation}
R^{M}_{A}(g)=R^{M}_{A}(hg, h)|_{h=e}\,,\quad
\overline{R}^{A}_{M}(g)=\overline{L}^{A}_{M}(h, hg)|_{h=e}\,,
\label{a110a}
\end{equation}

Here, as in section (\ref{s1}) the first lot of Latin indices $A, B,
\dots =\overline{1, p}$ is used to indicate the group degrees of
freedom (i.e. the frame of $T_{e}G$), and the final one $M, N,
\dots=\overline{1, p}$ describes the tensor degrees of freedom.

The structure equations for the left Lie algebra is:

\begin{equation}
[L_{A}, L_{B}]=c^{C}{}_{AB} L_{C}\,,
\label{a113}
\end{equation}
or, in a local form

\begin{equation}
L^{M}_{A}\partial_{M} L^{N}_{B}-L^{M}_{B}\partial_{M} L^{N}_{A}=
c^{C}{}_{AB} L^{N}_{C}\,.
\label{a114}
\end{equation}

The right Lie algebra has the basis $\{R_{A}=R^{M}_{A}\partial_{M}\}$
and obeys the similar structure equations, that can be obtained by the
following replacement

$$
L^{M}_{A}\longrightarrow  R^{M}_{A}\,,\quad
c^{C}{}_{AB}\longrightarrow  -c^{C}{}_{AB}\,.
$$

The inverse matrix $\overline{L}$ satisfies the Maurer-Cartan equations

\begin{equation}
\partial_{M}\overline{L}^{A}_{N}-\partial_{N}\overline{L}^{A}_{M}=
-c^{A}{}_{BC}\overline{L}^{B}_{M} \overline{L}^{C}_{N}\,.
\label{a115}
\end{equation}
Note, that left and right Lie algebras are commutative

\begin{equation}
R^{M}_{A}\partial_{M} L^{N}_{B}=L^{M}_{B}\partial_{M} R^{N}_{A}\,.
\label{a116}
\end{equation}

\subsection*{Action of Lie Group on Manifold}

Let $M$ and $G$ be a smooth manifold and the Lie group of right
transformations on $M$ defined by the map

\begin{equation}
\begin{array}{rcl}
\rho:&M\times G&\longrightarrow  M\\
     &   (x, g)&\longrightarrow  y:=\rho_{g} x\,,
\end{array}
\label{a118}
\end{equation}
with the following properties

\begin{enumerate}
\item
$\forall g_{1, 2}\in G$, $\rho_{g_{1}}\circ
\rho_{g_{2}}=\rho_{g_{2}g_{1}}$;

\item
$\rho_{e}={\rm id}_{M}$ (the identity map);

\item
$\forall g\in G$, $\rho_{g}: M\to M$ is a diffeomorphism ($\rho_{g}\in
\diff(M)$).
\end{enumerate}

The realization (\ref{a118}) of the Lie group as the
transformation group induces the realization of the Lie algebra by the
following procedure.

Let $A\in\lie(G)$, $g_{A}(\tau):=\exp (\tau A)\in G$ be an integral
curve of $A$ (one-parametric subgroup of $G$, $g_{A}(0)=e$).
Then the vector

$$
\tilde{\rho}_{A}|_{x}=
\left. \frac{d}{d\tau}\right|_{\tau=0} \rho_{\exp (\tau A)}(x)
\in T_{x} M
$$
defines the realization of $A\in \lie (G)$.
Due to the homomorphic nature of the map $\rho:G\to\diff(M)$,
the vector subspace $\tilde{\rho}_{\lie (G)}\subset\vect(M)$
is finite dimensional ($\vect(M)$ denote the space of vector
fields over $M$).

A Lie group can be realized on itself with the use of left [right]
translations. The induced realization of the Lie algebra coincides with
the Lie algebra of right [left] invariant vector fields.
The basic element of $T_{e}G$

$$
T_{A}|_{e}=\delta^{M}_{A} \partial_{M}|_{e}
$$
is represented by a vector

\begin{equation}
\left.\frac{d}{d\tau}\right|_{\tau=0}
x^{M}(g g_{A}(\tau))\partial_{M}|_{g}=
L^{M}_{A}(g)\partial_{M}|_{g}\,.
\label{a122}
\end{equation}
Similarly, for the left action we have

\begin{equation}
\left.\frac{d}{d\tau}\right|_{\tau=0}
x^{M}(g_{A}(\tau) g)\partial_{M}|_{g}=
R^{M}_{A}(g)\partial_{M}|_{g}\,.
\label{a123}
\end{equation}

\subsection*{Metric of Lie Group}

The tangent space $T_{e}G$ at the unit element $e\in G$ is a
$p$-dimensional vector space. We assume that there exists a scalar
product defined by

\begin{equation}
\begin{array}{rcl}
(\cdot\, , \cdot):&T_{e}G\times T_{e}G& \longrightarrow  {\bf R}\\
& \left\langle A|_{e}, B|_{e} \right\rangle&\longrightarrow  (A|_{e}, B|_{e})\,.
\end{array}
\label{a124}
\end{equation}
The scalar products of the basic elements $\{T_{A}|_{e}:
A=\overline{1, p}\}$ form the matrix

\begin{equation}
\eta_{AB}:=(T_{A}|_{e}, T_{B}|_{e})={\rm const}
\label{a125}
\end{equation}
that obeys the tensor transformation law under $Gl_{p}$
transformations in $T_{e}G$.

Using (\ref{a125}) one can easily show that the scalar product of
left-invariant fields $\{L_{A}\}$ on $G$ equals to the matrix
(\ref{a125}):

\begin{equation}
(L_{A}|_{g}, L_{B}|_{g})=(L_{A}|_{e}, L_{B}|_{e})\equiv
(T_{A}|_{e}, T_{B}|_{e})=\eta_{AB}\,.
\label{a126}
\end{equation}

This scalar product can be identified with the left-invariant metric of
$G$. If $G$ admits a subgroup $H\subset G$, the metric can be chosen
in the diagonal form

\begin{equation}
\{\eta_{AB}\}=
\left(
\begin{array}{cc}
g_{ab}&0\\
0&g_{ij}
\end{array}
\right)\,,
\quad
\{\eta^{AB}\}=
\left(
\begin{array}{cc}
g^{ab}&0\\
0&g^{ij}
\end{array}
\right)\,,
\label{a127}
\end{equation}
by means of $Gl_{p}$ transformation. The tensor $\{g_{ij}\}$
in (\ref{a127}) has the sense of the metric of $T_{e} H$.

The components of the metric of $G$ can be defined with respect to the
holonomic frame of $\vect(G)$ as the following scalar product

\begin{equation}
\eta_{MN}(g):=
\left(  \left. \frac{\partial }{\partial x^{M}}\right|_{g},
\left. \frac{\partial }{\partial x^{N}}\right|_{g} \right)\,.
\label{a127a}
\end{equation}
Thus, using (\ref{a126}), (\ref{a127a}) we can write

$$
\eta_{AB}=\left( L_{A}|_{g}, L_{B}|_{g} \right)=
L^{M}_{A}(g) L^{N}_{B}(g)
\left(  \left. \frac{\partial }{\partial x^{M}}\right|_{g},
\left. \frac{\partial }{\partial x^{N}}\right|_{g} \right)\,.
$$
Therefore,

\begin{equation}
\eta_{MN}=\overline{L}^{A}_{M}(g)\,\overline{L}^{B}_{N}(g)\eta_{AB}
\label{a128}
\end{equation}
is the left-invariant metric of $G$.

Under the coordinate transformation being performed by the right
translations the metric (\ref{a128}) transforms as

\begin{equation}
\eta_{M_{1}M_{2}}(g_{1}g_{2})=
\frac{\partial x^{N_{1}}(g_{2})}{\partial x^{M_{1}}(g_{1}g_{2})}
\frac{\partial x^{N_{2}}(g_{2})}{\partial x^{M_{2}}(g_{1}g_{2})}
\eta_{N_{1}N_{2}}(g_{2})\,.
\label{a130}
\end{equation}
Such a transformation law means that the right translations on $G$ are
isometric transformations of the metric (\ref{a128}). Equivalently, the
Lie variation of (\ref{a128}) associated with the right-invariant
vector field vanishes, i.e. the right Lie algebra consists of Killing
vectors.

However, the left-invariant vector field is not a Killing vector of
(\ref{a128}) in general case. It is possible if and only if the Lie
group is semisimple. In this case $\{\eta_{AB}\}$ coincide with an
adjoint-invariant Killing-Cartan form and the right-invariant metric of
$G$ is the same as the left-invariant one. Therefore, Killing vectors
of $\{\eta_{MN}\}$ correspond to both Lie algebras.

\section*{Appendix 2. Connection 1-form of Principal Fiber Bundle}
\setcounter{equation}{0}
\renewcommand{\theequation}{A2.\arabic{equation}}

The connection 1-form $\omega\in {\cal A}^{1}(G, \lie(H))$ (see, for
example, \cite{kobayashi}) of the principal fiber bundle $G(G/H, H)$
can be defined as the restriction of the Maurer-Cartan form of $G$
to $H$ by the identification $\lie(H)$ with $T_{e} H$:

\begin{equation}
\omega: \vect(G)\longrightarrow  \lie(H)\cong T_{e} H\,.
\label{a21}
\end{equation}
In this case we can write

\begin{equation}
\omega|_{g}=\omega_{M}(g) d\,x^{M}|_{g}\,,\quad
\omega_{M}(g)=\overline{L}^{i}_{M} T_{i}|_{e}\,,
\label{a22}
\end{equation}
where $\{T_{i}|_{e}: i=\overline{n+1, p}\}$ denotes a basis of
$T_{e} H$.

In the local coordinate system corresponded to $(m+n)$-decomposition,
(where $L^{\alpha}_{i}=0$), the connection 1-form has the following form

\begin{equation}
\omega|_{G}=\omega_{M} d\,x^{M}|_{g}=
\overline{L}^{i}_{M} L^{N}_{i}
d\,x^{M}|_{g}\otimes\left. \frac{\partial }{\partial x^{N}}\right|_{g}=
\omega^{\mu}_{M} d\,x^{M}|_{g}\otimes
\left. \frac{\partial }{\partial x^{\mu}}\right|_{g}\,.
\label{a23}
\end{equation}

Due to the properties of the matrices $L$ and $\overline{L}$
in such a coordinate system one can observe that

\begin{equation}
\omega^{\mu}=\omega^{\mu}_{M} d\,x^{M}=
d\,x^{\mu}+\overline{L}^{i}_{\alpha} L^{\mu}_{i} d\,x^{\alpha}=
d\,x^{\mu}+A^{\mu}_{\alpha} d\,x^{\alpha}\,,
\label{a24}
\end{equation}
where $A^{\mu}_{i}=\overline{L}^{i}_{\alpha} L^{\mu}_{i}$ has the sense
of a gauge field.


\begin{thebibliography}{99}

\bibitem{schwinger} J.Schwinger, {\it Phys.Rev.} {\bf 82}, 914 (1951);
 {\it ibid.} {\bf 91}, 713 (1953);\newline
see also: D. V. Volkov and S. V. Peletminsky
{\it JETP} {\bf 37}, 170 (1959);\newline
N. Ogawa et al., {\it Prog. Theor. Phys.} {\bf 96}, 437 (1996).

\bibitem{ourpaper} submitted to {\it Europ. J. Phys.}

\bibitem{kaluza} See, for example:
                 {\it Fortschr. der Phys.} {\bf 32}, 607 (1984)

\bibitem{tsutsui}
 D. McMullan and I. Tsutsui, {\it Ann. Phys.} {\bf 237}, 269 (1995);
\newline
 Y. Ohnuki and S. Kitakado {\it J. Math. Phys.} {\bf 34}, 2827 (1993).

\bibitem{kobayashi} S. Kobayashi, Nomidzu,
                    {\it Foundations of Differential Geometry}, vol.II

\bibitem{hadron}
K. Fujii, K.-I. Sato, N. Toyota, A.P. Kobushkin,
{\it Phys. Rev. Let., 1987}, {\bf 58}, 651 (1987).
\newline
K. Fujii, A.P. Kobushkin, K.-I. Sato, N. Toyota,
{\it Phys. Rev.} D {\bf 35}, 1896 (1987).
\newline
K. Fujii, K.-I. Sato, N. Toyota,
{\it Phys. Rev.} D {\bf 37}, 3663 (1987).
\newline
K. Fujii, N. Ogawa, K.-I. Sato, N.M. Chepilko, A.P. Kobushkin,
T. Okazaki, {\it Phys. Rev.} D {\bf 44}, 3237 (1991).

\bibitem{groups} A. Kirillov, {\it Elements of Representation Theory},
Springer-Verlag, 1976

\end{thebibliography}
\end{document}